# Behavior of Solar Cycles 23 and 24 Revealed by Microwave Observations

N. Gopalswamy[1], S. Yashiro[1,2], P. Mäkelä[1,2], G. Michalek[3], K. Shibasaki[4], and D. H. Hathaway[5]

[1]NASA Goddard Space Flight Center, Greenbelt, Maryland

[2]The Catholic University of America, Washington, DC

[3]Jagielloninan University, Krakow, Poland

[4]Nobeyama Solar Radio Observatory, Nobeyama, Japan

[5]NASA Marshall Space Flight Center, Huntsville, Alabama

ABSTRACT

Using magnetic and microwave butterfly diagrams, we compare the behavior of solar polar regions to show that (i) the polar magnetic field and the microwave brightness temperature during the solar minimum substantially diminished during the cycle 23/24 minimum compared to the 22/23 minimum. (ii) The polar microwave brightness temperature (b) seems to be a good proxy for the underlying magnetic field strength (B). The analysis indicates a relationship, $B = 0.0067 T_b - 70$, where B is in G and $T_b$ in K. (iii) Both the brightness temperature and the magnetic field strength show north-south asymmetry most of the time except for a short period during the maximum phase. (iv) The rush-to-the-pole phenomenon observed in the prominence eruption activity seems to be complete in the northern hemisphere as of March 2012. (v) The decline of the microwave brightness temperature in the north polar region to the quiet-Sun levels and the sustained prominence eruption activity poleward of $60^o$N suggest that solar maximum conditions have arrived at the northern hemisphere. The southern hemisphere continues to exhibit conditions corresponding to the rise phase of solar cycle 24.

## 1. Introduction

The peculiar solar minimum that followed the solar cycle 23 and the delayed onset of cycle 24 have received extensive attention (see the articles in Cranmer, Hoeksema, and Kohl, 2010). The delayed start of solar cycle 24 has also raised the question about its amplitude and when the cycle will reach its maximum. Some of the main characteristics that have been used to characterize the peculiar solar minimum are the reduced polar field strength, extremely low levels of activity and the long duration. These characteristics have been extensively studied mainly using photospheric and coronal observations. In particular, the polar field strength has been studied extensively using photospheric magnetograms. Microwave imaging observations at frequencies above 15 GHz provide a unique opportunity to study the behavior of the polar regions because coronal holes appear bright at these frequencies (see e.g., Kosugi, Ishiguro, and Shibasaki, 1986; Gopalswamy et al. 1999a,b; Nindos et al., 1999; Shibasaki, Alissandrakis, and Pohjolainen, 2011; Prosovetsky & Myagkova, 2011). The quiet-Sun microwave emission originates from the chromospheric layers, so we can study the polar regions at the chromospheric layer and compare it with photospheric observations.

Microwave imaging can also be used to study prominence eruptions (PEs), whose locations on the Sun also provide important information on various phases of the solar cycle such as the rush-to-the-pole phenomenon and the cessation of high-latitude activity that marks the time of polarity reversal (Gopalswamy et al. 2003a; Shimojo et al., 2006). In addition the near-Sun propagation of PEs and the associated coronal mass ejections (CMEs) also exhibit interesting interaction with the polar magnetic field, resulting in a systematic offset between the position angles of PEs and the associated CMEs (Gopalswamy and Thompson, 2000; Gopalswamy et al. 2003a; Cremades, Bothmer, and Tripathi, 2006; Gopalswamy et al. 2009; Lugaz et al. 2011; Panasenco et al. 2011).

In this paper, we make use of the long-term data base of microwave imaging observations at 17 GHz made by the Nobeyama radioheliograph (NoRH, Nakajima et al., 1994) to study the behavior of PEs and polar coronal holes over three solar cycles (22, 23, and 24). The microwave observations are available for cycles 22 (declining phase), 23 (full), and 24 (rise phase). Therefore, we can compare the 22/23 and 23/24 minima. Using the synoptic maps made from the daily 17 GHz microwave images, we construct the microwave butterfly diagram and compare it with the magnetic butterfly diagram to understand the low and high latitude characteristics of the three solar cycles. In particular, the microwave butterfly diagram is compared with the magnetic butterfly diagram to see how the reduction in polar field strength is reflected in the microwave brightness.

## 2. The Microwave Butterfly Diagram

Radio synoptic maps at 17 GHz are constructed from daily full-disk brightness temperature maps by the Nobeyama Radioheliograph. The typical spatial resolution of the microwave maps is 10 arcsec. The daily best images are made at local noon and the images are almost continuously available except when there is heavy rain and wet snow pileup. A central strip covering +/- 20 degrees from the central meridian are cut from the daily images and assembled on the Carrington coordinate grid, as is done for other types of synoptic charts (Shibasaki, 1998; Gopalswamy, Thompson, and Shibasaki, 1998). The synoptic map for each rotation is averaged in the longitudinal direction to obtain a single column of brightness temperature values. These columns are then assembled to obtain the microwave butterfly diagram (Gelfreikh et al., 2002). There are 263 Carrington rotations (CR#1858 to #2120) from June 1992 to March 2012, covering the decay phase of the solar cycle 22, the complete cycle 23, and the rise phase of cycle 24.

Figure 1 shows the NoRH microwave butterfly diagram and its comparison with the magnetic butterfly diagram. The data for the magnetic butterfly diagram came from three sources – the Kitt Peak National Observatory (KPNO) synoptic maps, the Solar and Heliospheric Observatory (SOHO) mission's Michelson Doppler Imager (MDI) synoptic maps, and the maps from the Synoptic Optical Long-term Investigations of the Sun (SOLIS) facility. The KPNO synoptic maps have a size of 360 x180 in longitude and *sin* (latitude). The other two maps were smoothed and resampled to this size. The data from these maps were averaged in longitude to get the butterfly diagram.

The microwave butterfly diagram contains 263 rotations corresponding to the period of operation of NoRH. Two prominent features can be seen on the synoptic maps. (1) The low-latitude brightness features in the northern and southern hemispheres corresponding to the sunspot activity. These are the active region belts starting around $40^o$ in the beginning of the cycle and ending at the equator towards the end of the sunspot cycle. (2) The high-latitude brightness enhancement peaking between cycles. The brightness enhancement happens in polar coronal holes, where the poloidal magnetic field of the Sun peaks during solar minima. During solar maxima, the polar coronal holes disappear and the microwave brightness returns to quiet Sun level (about 10,000 K). Thus the microwave butterfly diagram contains information on both toroidal and poloidal magnetic fields of the Sun and their variation as a function of the phase of the solar cycle.

## 3. Correlation between Microwave Brightness Temperature and Polar Magnetic Field Strength

The photospheric layers underlying coronal holes can be distinguished by unipolar magnetic fields, enhanced with respect to the quiet Sun. Therefore, one would expect a connection to the microwave emission originating from the overlying chromospheric layers. Gopalswamy, Shibasaki, and Salem (2000) investigated a set of 71 equatorial coronal holes and found that the microwave brightness temperature is correlated with the peak magnitude of the photospheric magnetic field underlying the coronal hole. Polar coronal holes are closely related to solar minima when the polar field strength is high (see, e.g., Benevolenskaya, 2010), so we expect them to have similar relationship between brightness temperature and magnetic field strength. We now compare the microwave and magnetic butterfly diagrams to establish a quantitative relation between the microwave brightness temperature and the polar field strength. We average the brightness temperature (Tb) and field strength (B) poleward of $60^{o}$ latitudes in the respective butterfly diagrams to get the average Tb and B for each pole separately. We then perform rotation by rotation comparison of Tb and B to assess how closely the two are related. Figure 2 shows the evolution of B from 1992 to the beginning of 2012. The sunspot number (SSN) is overlaid to show the phases of the solar cycle. The field strength starts rising from the solar maximum of cycle 22 and peaks during the 22/23 minimum around 1996. During the maximum phase of cycle 23, the poles reverse their polarity and the polar field strength increases again. Clearly, the polar field strengths never rise up to the level of cycle 22/23 minimum as has been also reported by many others (see e.g., Hathaway, 2010; Petrie, 2012). Finally, the polar field strength has started to decline with the rise of solar cycle 24 and is currently at zero level.

In the bottom part of Figure 2, we see that the evolution of Tb closely follows that of B at both poles. The difference between the north and south polar regions is reflected in both quantities. Most of the time, Tb and B are higher in the south pole compared to the corresponding values in the north. The brief dominance of the north pole during 2001-2004 is also similar in Tb and B. The average values of Tb and B during the cycle 22/23 and 23/24 minima are showed by the horizontal lines. B declines by 3.5 G from the 22/23 minimum to the 23/24 minimum. Tb declines by 200 K. Both Tb and B show north-south asymmetry, with south dominating throughout except around the time of solar maximum when the north polar region had slightly higher values. Solar maximum conditions seem to have arrived to the north polar region in the beginning of 2012 because Tb has declined to the quiet Sun level (similar to the level in the beginning of year 2000 during cycle 23 maximum). However, Tb in the south polar region is still quite high suggesting that the rise-phase conditions continue to prevail in the south. Such an offset between the north and south can also be seen during the cycle 23 maximum.

Figure 3 compares Tb and B for each Carrington Rotation, showing the north-south asymmetry and the difference between the cycle 22/23 and 23/24 minima. Tb and B show high correlation in the south polar region (correlation coefficient, CC = 0.86 for 22/23 minimum and 0.71 for the 23/24 minimum) and a moderate one in the north polar region (CC =0.64 for cycle 22/23 minimum and 0.54 for the 23/24 minimum). Using only those Carrington rotations when the B0 angle is in the range ±1° did not significantly improve the correlation. However, there was significant improvement when we used only those rotations for which $|B0| \geq 6°.25$ (when one of the poles was visible to an Earth observer) with the following CC: 0.60 (North, cycle 22/23), 0.89 (North, cycle 23/24), 0.92 (South, cycle 22/23), and 0.84 (South, cycle 23/24). When all the Tb and B data were combined, CC becomes 0.86 with a regression line given by B =

$(6.7\pm0.4)10^{-3}$ Tb $-(70\pm5)$. Note that the microwave Tb corresponds to the upper chromospheric layer, which is significantly different from the photosphere. The less than perfect correlation between Tb and B may be due this difference, but this needs further investigation. In the south, the polarity remained negative (south-pointing) until the reversal during the middle of 2002. In the north, the polarity was positive until the end of 2000, when the polarity reversed and became negative. The small Tb and B values correspond to solar maximum conditions when the polar coronal holes disappeared. In the south, Tb was lower during 22/23 minimum compared to the 23/24 minimum by ~450 K. The same is true in the north, except the difference is smaller (150 K). The north-south asymmetry was present throughout the interval studied except for seven rotations during the maximum phase of cycle 23. The south polar Tb and B occupied a narrow range of values during 23/24 minimum compared to the larger ranges during the 22/23 minimum.

## 4. Prominence Eruption Activity

There is a high degree of association between prominence eruptions and CMEs (Munro et al., 1979; Hori and Culhane, 2002; Gopalswamy et al., 2003b). PEs are automatically detected in the 17 GHz images archived at the Nobeyama Radio Observatory (http://solar.nro.nao.ac.jp/norh/html/prominence/). The latitudes of PEs superposed on the microwave butterfly diagram in Fig. 4 (top) demonstrate the solar cycle variation of the eruption locations. In particular one can see the high latitude eruptions only during the solar maximum phase. Cessation of these high-latitude eruptions mark the end of the solar maximum and the Sun reverses the polarity of the global field. The sparse distribution of data points around the 23/24 minimum indicates that occurrence rate of the PEs is much smaller during the 23/24 minimum compared that during the 22/23 minimum. If we count all PEs in the rise phase, we find a PE rate of 1.4 per month for the 22/23 minimum, which is nearly a factor of 3 higher than the 0.5 PEs per month for the 23/24 minimum. It must be noted that the Nobeyama radioheliograph observes the Sun only for ~8 h per day, so all the PEs in a given day are not included. However, the number of eruptions is sufficiently large to show the solar cycle variation.

The position-angle correspondence between PEs and CMEs has also solar-cycle dependence: during solar minima, the PEs start at higher latitude and the corresponding CMEs appear at lower latitudes. This positive PE-CME offset is also thought to be an indicator of the strong polar fields (Gopalswamy and Thompson, 2000; Gopalswamy et al. 2003a; Cremades, Bothmer, and Tripathi, 2006; Gopalswamy et al. 2009; Lugaz et al. 2011; Panasenco et al. 2011). The offsets were computed from Nobeyama PE observations and SOHO CME observations. The intervals of positive PE-CME offset are delineated by the vertical lines for cycle 24 (see the bottom portion of Fig. 4). For cycle 23, the CME identification was occasionally not possible either due to data gap or the lack of association. Before 1996, there were no coronagraphic observations to compare with the Nobeyama data. The offset intervals extend from the second half of the minimum phase to the rise phase of the next cycle. During the first half of the solar minimum phase, the PEs occur only at lower latitudes, so the influence of the coronal hole is not felt by them, so there is no positive offset. The interval of positive offset is rather extended during the cycle 23/24 minimum. The average PE-CME offset for the 23/24 minimum is $19^{\circ}$, which is nearly the same as the offset during the 22/23 minimum ($18^{\circ}$). The polar field strength was diminished during the 23/24 minimum, but probably was sufficient to deflect the PEs. We also notice a strong north-south asymmetry in the number of PEs during the positive PE-CME offset interval: more PEs came from the north during both the minima.

The rush-to-the-pole (RTTP) behavior of filaments and prominences are known for a long time and the completion of the phenomenon happens during solar maximum (Hyder, 1965; McIntosh, 2003). RTTP can also be seen in the synoptic map of local intensity maxima in the FeIV line emission (Altrock, 2003; 2010). Gopalswamy et al. (2003a,b) found that the locations of PEs and the associated CMEs also spread to higher latitudes, typically at the rate of 10-12 degrees per year. The RTTP phenomenon and the cessation of high-latitude eruptions mark the time of polarity reversal. They also noted that when the PEs occur at latitudes >60$^o$, the tilt angle reaches its maximum value indicating solar maximum conditions. When we look at cycle 24, the PE locations have started crossing the 60$^o$ line in the beginning of year 2011. The sustained high-latitude PE activity coincides with the significant decline of the polar microwave brightness, as during the cycle 23 maximum. Thus the PE activity indicates the completion of RTTP in the north and hence supports the prevalence of solar maximum conditions in the north polar region. The PE activity in the southern hemisphere has still not crossed the 60$^o$ line, so the solar maximum conditions have not arrived in the southern hemisphere. Although a recent report by Altrock (2010) did not see RTTP in the south, the PEs do indicate the onset in the middle of year 2009 around 40oS. Such delay in the southern hemisphere was also noticed in previous cycles, which leads to the delayed polarity reversal at the south pole. The high-latitude PE and CME activities persist for 2-3 years after crossing 60$^o$ latitude and when the activity ends, the polarity reverses. It appears that the polarity reversal during the cycle 24 maximum will have a north-south asymmetry similar to that of the cycle 23 maximum.

## 5. Summary and Conclusions

We constructed the microwave butterfly diagram using 17 GHz images obtained by the Nobeyama radioheliograph for the period June 1992 to March 2012 spanning Carrington rotations 1858 to 2120. The microwave butterfly diagram provides information on the high and low latitude properties of the solar activity cycle and compares well with the magnetic butterfly diagram. We also used the rate and latitudes of prominence eruptions to study their variations with the phase of the solar cycle. We also computed the position-angle offset between prominence eruptions and the associated CMEs and confirmed that the positive-only offset occurs during the very beginning of the solar activity cycle. The primary conclusions of this investigation are:

(i) The microwave butterfly diagram reveals many features of the solar activity cycle at low and high latitudes. The microwave emission originates from the upper chromosphere where the temperature is around 10,000 K and can be considered as the chromospheric counterpart of the magnetic butterfly diagram constructed from photospheric magnetograms.

(ii) The microwave brightness temperature during the 23/24 minimum is substantially diminished compared to the 22/23 minimum, consistent with the decrease in the polar magnetic field strength.

(iii) The polar magnetic field strength and the microwave brightness temperature are reasonably correlated suggesting that the latter is a good indicator of the underlying magnetic field strength: $B = 0.0067 T_b - 70$, where B is in G and $T_b$ is in K.

(iv) Both the brightness temperature and the magnetic field strength show north-south asymmetry, with the south pole dominating most of the time except for a short time during the maximum phase of solar cycle 23.

(v) The positive offset between the position angles of prominence eruptions and the associated CMEs, first identified during solar cycle 23 has been confirmed during cycle 24. The duration of solely positive offset is more extended during cycle 24. However, the average positive offset seems to be similar between the two cycles.

(vi) The rush-to-the-pole phenomenon observed in the prominence eruption seems to be complete in the northern hemisphere. It has started in the southern hemisphere, but has not crossed the 60$^o$ latitude.

(vii) The decline of the microwave brightness temperature in the north polar region to the quiet sun values and the sustained prominence eruption activity poleward of 60$^o$N suggest that solar maximum conditions prevail in the northern hemisphere with the polar field vanishing. The southern hemisphere, on the other hand, still has conditions corresponding to the rise phase of solar cycle 24.

NG and SY acknowledge the travel support during their visit to the Nobeyama Solar Radio Observatory, where part of this work was done. SOHO is a project of international cooperation between ESA and NASA. Work supported by NASA/LWS program.

Figure Captions

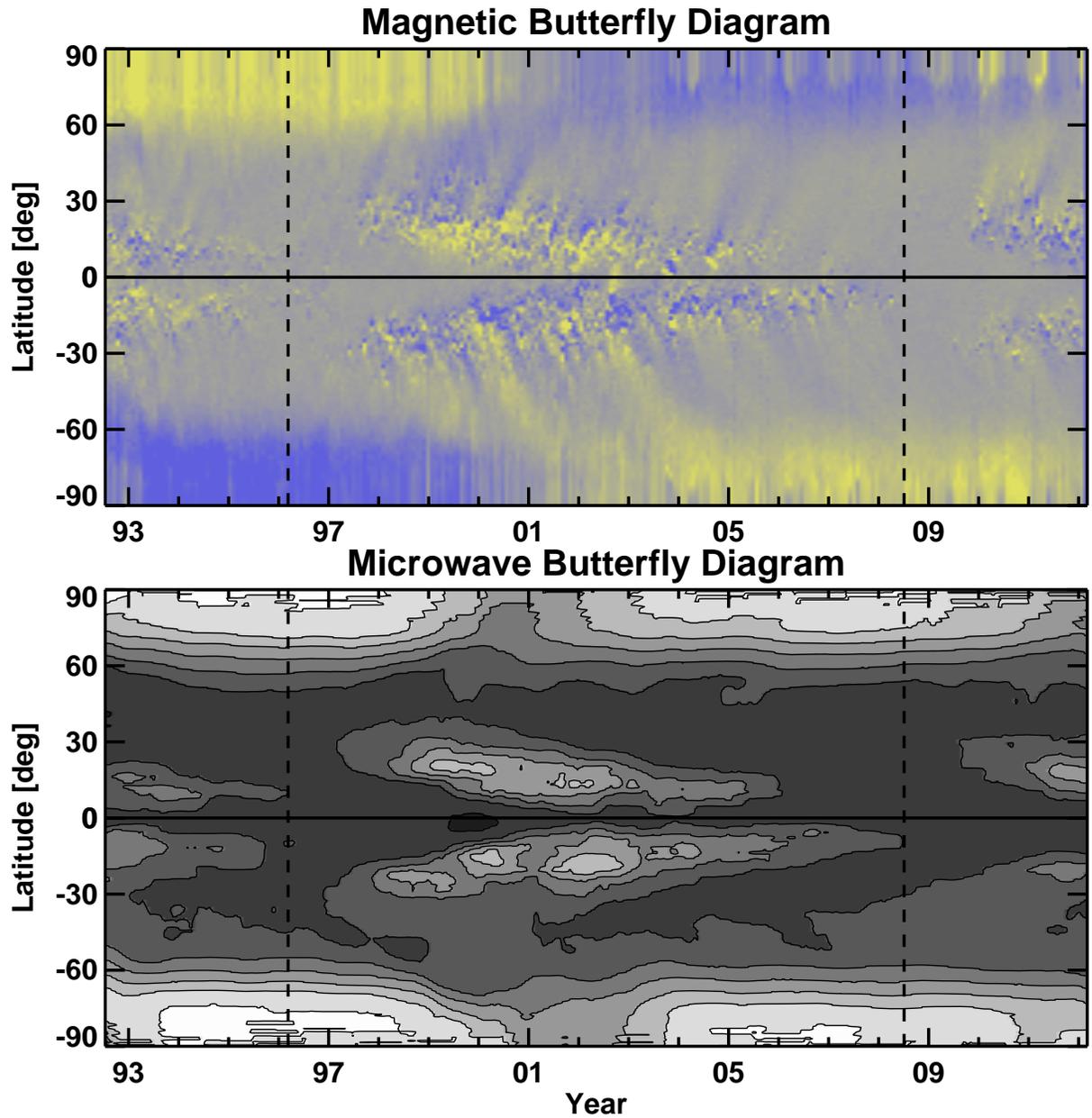

Figure 1. (top) The magnetic butterfly diagram from 1992 to the present constructed mainly from Kitt Peak National Observatory with SOHO/MDI data filling several gaps. Blue and yellow indicate positive and negative polarities, respectively. The magnetic field strength ranges from -10 G to + 10 G. (bottom) The microwave butterfly diagram constructed from the Nobeyama

radioheliograph images at 17 GHz. A 13-rotation smoothing has been used along the time axis to eliminate the periodic variation due to solar B0-angle variation. The contour levels are at 10000, 10300, 10609, 10927, 11255, 11592, and 11940 K. The vertical dashed lines mark the ends of cycles 22 and 23 around March 1996 and July 2008, respectively determined based on the low-latitude activity in microwaves.

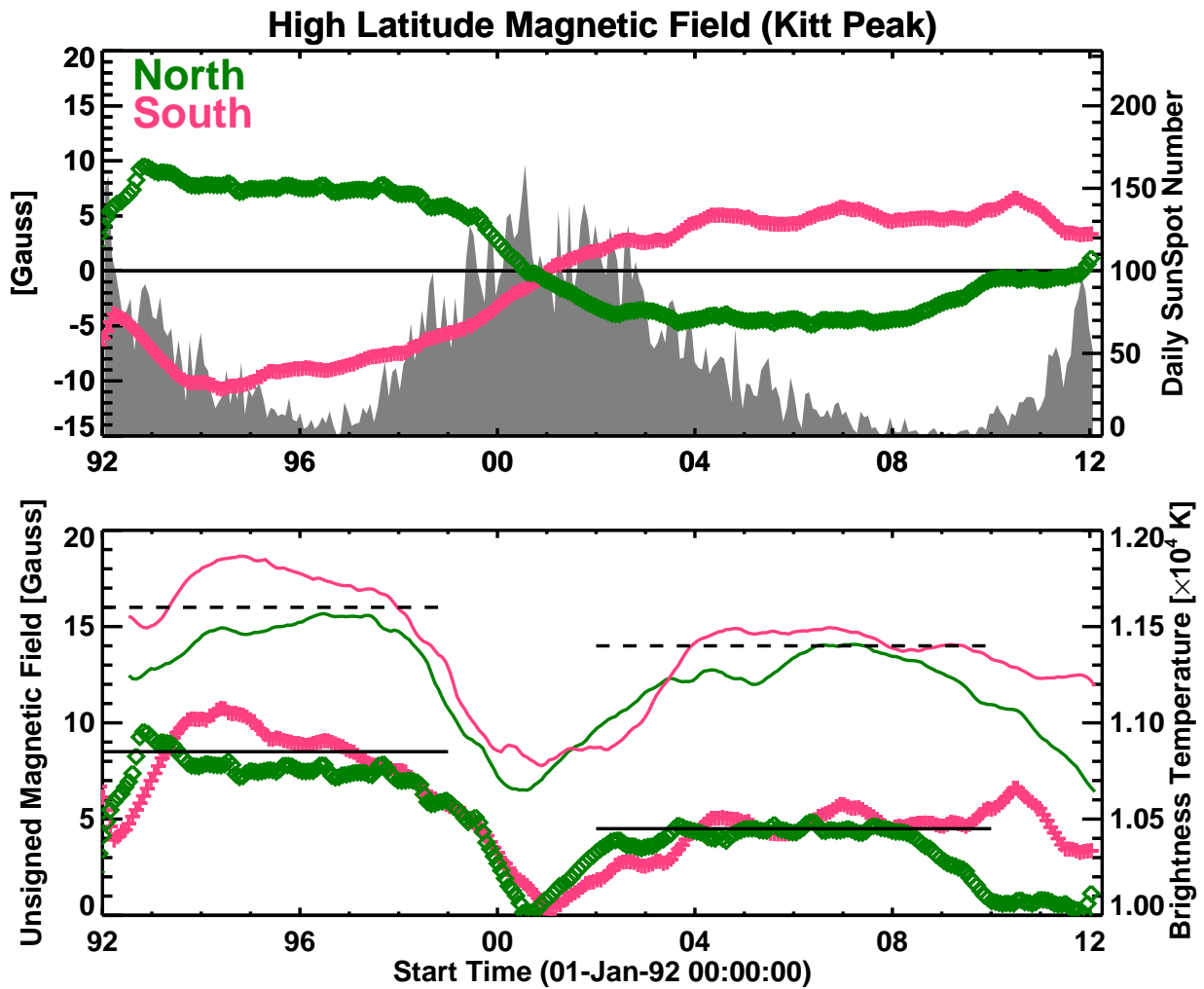

Figure 2. (top) The field strength in the north and south poles averaged over latitudes poleward of 60°. The sunspot number is plotted for reference. (bottom) The microwave brightness temperature also averaged over latitudes poleward of 60° shown separately for north and south

poles. The average magnitude of B values from the north and south poles are also plotted for reference. The horizontal dashed (solid) lines are for guidance to show the difference between the 22/23 and 23/24 minima.

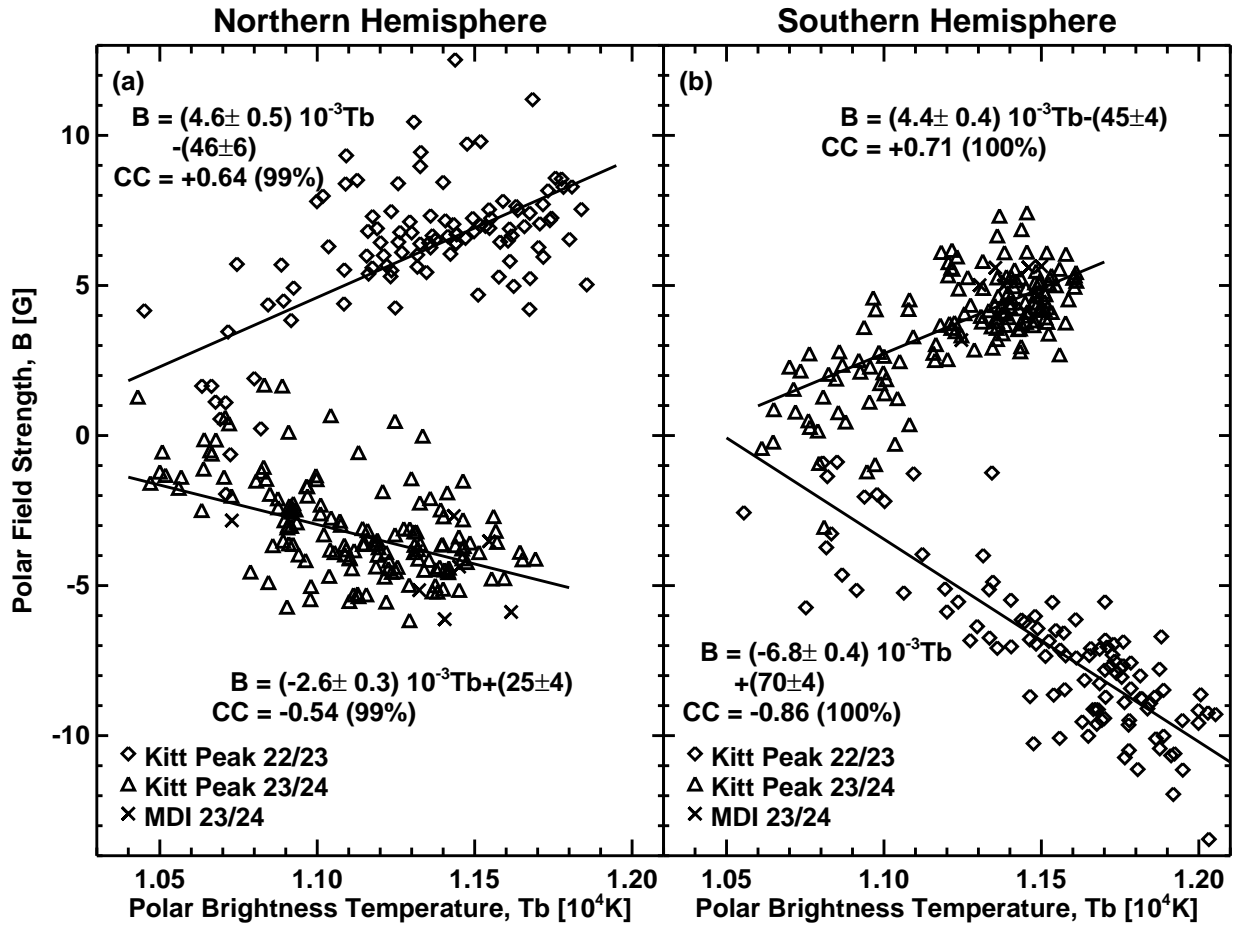

Figure 3. Scatter plot between B and Tb for the north (a) and south (b) polar regions. The positive (negative) values in the north polar region correspond to cycle 22/23 (23/24) minimum. The negative (positive) values in the south polar region corresponds to cycle 22/23 (23/24) minimum. The regression lines and the correlation coefficients (CC) are marked on the plots. The 1-sigma errors in the fit coefficients are given, which range from 11.5% to 15.2% in the northern hemisphere and 5.4% to 9.1% in the southern hemisphere. The numbers in parentheses next to CC is the probability that the correlation is not by chance.

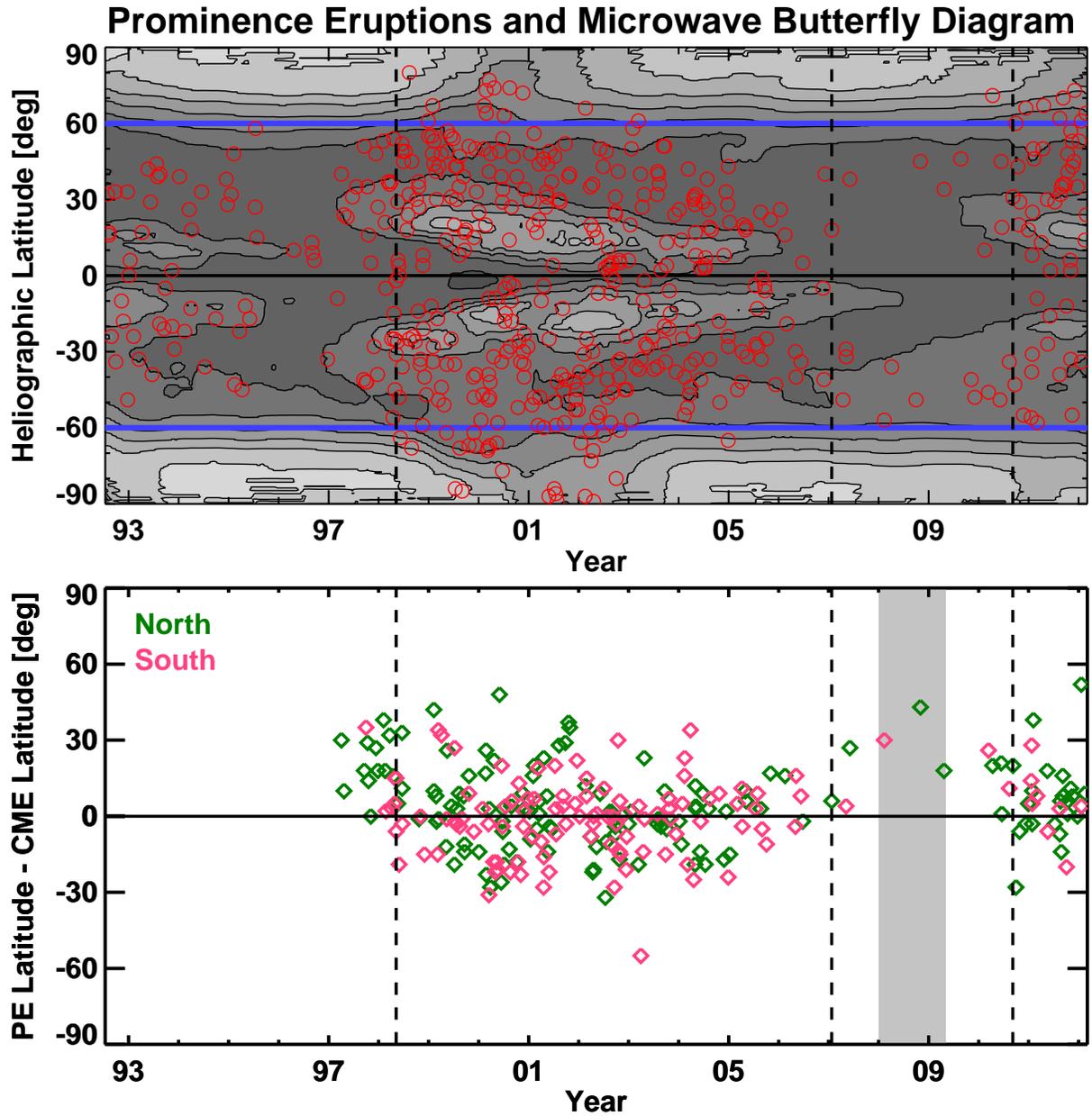

Figure 4. (top) Superposition of PE latitudes (open circles) on the microwave butterfly diagram. (bottom) The PE – CME latitude offset plotted as a function of time. The intervals in which the offset is exclusively positive are bracketed by the vertical dashed lines. The CME latitudes were obtained from SOHO's Large Angle and Spectrometric Coronagraph (LASCO) images. PEs occurring in the northern and southern hemispheres are distinguished by different colors. Note that this interval is rather extended during the 23/24 minimum, reflecting the extended nature of the microwave polar brightening. The vertical shaded rectangle marks the duration when active regions from cycles 23 and 24 were present, indicating the overlap between the two cycles. Although PEs were observed by the Nobeyama Radioheliograph before 1996, there were no coronagraphic observations, so we were not able to compute PE-CME offset.